\documentclass[preprint, prd,showpacs,nofootinbib]{revtex4}
\usepackage{dcolumn}
\usepackage{amsmath}
\usepackage{revsymb}
\usepackage{graphicx}
\usepackage{psfrag}

\newcommand{\be}{\begin{equation}}
\newcommand{\ee}{\end{equation}}
\newcommand{\eps}{\varepsilon}
\newcommand{\ga}{\alpha}
\newcommand{\gb}{\beta}
\newcommand{\gd}{\delta}
\newcommand{\gl}{\lambda}
\newcommand{\gs}{\sigma}

\newcommand{\rf}[1]{(\ref{#1})}
\newcommand{\mpl}[1]{M_{\text{Pl}}^{(#1)}}
\newcommand{\GeV}{\text{GeV}}
\def\tu{{\bar u}}
\def\tv{{\bar v}}
\def\tx{{\bar x}}
\def\tz{{\bar z}}
\def\Tt{{\bar t}}
\def\tr{{\bar r}}
\begin{document}
\preprint{hep-ph/0401116}
\preprint{ITFA-2004-02} 
\title{Black Hole Production in
  Particle Collisions \\ and Higher Curvature Gravity}
\author{Vyacheslav S.~Rychkov}
\email{rychkov@science.uva.nl}%
\affiliation{Insituut voor
   Theoretische Fysica, Universiteit van Amsterdam\\ Valckenierstraat
   65, 1018XE Amsterdam, The Netherlands}
\date{April 2004}
\begin{abstract}
The problem of black hole production in transplanckian
particle collisions is revisited, in the context of large extra dimensions
scenarios of TeV-scale gravity. The validity 
of the standard description of this process (two colliding 
Aichelburg-Sexl shock waves in classical
Einstein gravity) is questioned. It is observed that the classical
spacetime has large curvature along the transverse collision plane, as
signaled by the curvature invariant
$(R_{\mu\nu\lambda\sigma})^2$. Thus quantum gravity effects, and in
particular higher curvature corrections to the Einstein gravity,
cannot be ignored. To give a specific example of what may happen, the
collision is re-analyzed in the Einstein-Lanczos-Lovelock gravity theory,
which modifies the Einstein-Hilbert Lagrangian by
adding a particular `Gauss-Bonnet' combination of curvature squared terms. The analysis uses a series of approximations,
which reduce the field equations to a tractable second order nonlinear
PDE of the Monge-Amp\`ere type. It is found that the resulting spacetime
is significantly different from the pure Einstein case in the
future of the transverse collision plane. These considerations cast
serious doubts on the geometric cross section estimate, 
which is based on the classical Einstein gravity
description of the black hole production process.
\end{abstract}
\maketitle

\section{Introduction}

All present-day macroscopic experimental data are consistent with
gravity being described by the 4-dimensional Einstein-Hilbert action
\be
\label{4d}
S=-\frac 1{16\pi G} \int R\sqrt{-g}\,d^4x 
\ee
In the microworld no gravitational effects have been observed, and
indeed the 4-dimensional (4d) Newton constant $G$ in \rf{4d} is so small
that such effects would be totally negligible up to unreachable
collider energies of $\mpl{4}\sim 10^{19}\ \GeV$ (the 4d
Planck scale).

The idea of {\it large extra dimensions} allows to imagine a world in
which gravity is strong enough to play a role in elementary particle
collisions at accessible energies. 
In this world Einstein gravity propagates in $D>4$ dimensions, out of
which $D-4$ are curled up in a compact manifold of size $\ell\gg
1/\mpl{4}$. 
At distances $\gg \ell$ such gravity would be described by an
effective 4d action of the form \rf{4d}. To get the effective 4d Newton constant
right, we have to set the $D$-dimensional Planck scale at \cite{ADD}
\be
\mpl{D}\sim \mpl{4}(\mpl{4}\ell)^{-\frac{D-4}{D-2}}\ll \mpl{4}.
\ee
This means that such gravity will be much stronger than we are used to
believe at short distances $\ll\ell$.

Motivated by the hierarchy problem, Arkani-Hamed, Dimopoulos and Dvali \cite{ADD}
proposed to push this idea to the extreme and lower $M_{\text{Pl}}$
all the way down to the electroweak scale $M_{\text{EW}}\sim 1\
\text{TeV}$. This proposal requires extra dimensions ranging from a mm to a fermi
for $6\le D\le 11$ ($D=5$ leads to $\ell\sim 10^{13}\ \text{cm}$ and
is excluded). To make this consistent with the 4d Standard Model, we
have to assume that all the other fields but gravity do not feel the extra
dimensions. They must be confined to a 4d submanifold ({\it brane}) of
the $D$-dimensional world ({\it bulk}).

The simplest collider signature of such {braneworld} scenarios
would be apparent energy non-conservation due to produced gravitons
escaping into the bulk. Recent Tevatron searches \cite{Tevatron}
of such missing energy events found no statistically
significant effect and set a lower bound of $\sim 0.6\ \text{TeV}$ on
the $D$-dimensional Planck scale (for a general review of signals and
constraints, see \cite{Rev}). Of course the LHC with its c.m.~energy
of 14 TeV will be a much better probe of such phenomena.

In this paper I would like to contribute to the unfinished discussion
of another possible signature of TeV-scale gravity --- black hole
production in transplanckian collisions (see reviews \cite{Landsberg}).

 The common current opinion is that this process may be adequately
described using classical general relativity, and that a single large 
black hole will form for a range of impact parameters (in the $D$-dimensional Planck units)
\be
\label{range}
b\lesssim R_{\text{Schw}}(E)\sim E^{1/(D-3)}.
\ee

Here $E\gg 1$ is the energy of colliding particles, and
$R_{\text{Schw}}(E)$ is the Schwarzschild radius of the
$D$-dimensional static black hole of mass $E$.
If production cross section 
\be
\label{geom}
\sigma\sim\pi\bigl[ R_{\text{Schw}}(E)\bigr]^2
\ee
based on this estimate (known as the {\it
  geometric cross section}) were true, the LHC
would produce black holes at a rate $\sim$1 Hz for $M_{\text{Pl}}=1$
TeV, becoming a black hole factory \cite{factory}.

In older studies of transplanckian collisions \cite{Ven} 
the authors usually carefully stated that in the range 
\rf{range} a black hole is {\it expected} to form. 
Indeed, the eikonal expansion, normally used to analyze quantum gravity
in the transplanckian regime, breaks down exactly in that range.
It seems that the geometric cross section proposal was first put
forward unequivocally by Banks and Fischler \cite{BF}.

Later on, an appealing theoretical argument backing \rf{range}, \rf{geom} was proposed by 
Eardley and Giddings \cite{EG}.
In the first step of the argument the problem of black hole formation
is analyzed in classical Einstein gravity using the {\it closed
  trapped surface} (CTS) method. A well-defined mathematical problem
of finding a CTS in the spacetime formed by superposing two
Aichelburg-Sexl shock waves is formulated and solved. For the range of
impact parameters when a CTS is found, black hole formation is
concluded by invoking the Cosmic Censorship Conjecture.

In the second step the aim is to argue that quantum corrections to
general relativity are not likely to modify the conclusions, because
the classical spacetime has small curvature in the regions relevant
for the trapped surface evolution. However, it is this second step where I
disagree with Eardley and Giddings.\footnote{It should be mentioned that the validity of the geometric
  cross section was also challenged by Voloshin \cite{Vol}. Some
  of his criticisms were addressed in \cite{EG,disc}, others still
  remain unanswered.
 There does not seem to be an obvious connection between Voloshin's and my reasons for critique.} 

As we will see, curvature becomes large on the transverse plane at
the moment when the particles pass each other, and in the
future of this plane corrections to Einstein gravity cannot be
ignored. In particular, if higher order curvature terms are present in
the effective gravitational Lagrangian (and we have no reason to
believe that they don't), they will become important precisely at this
moment. 

To see a specific example of how this might happen, we will add to the
Einstein-Hilbert Lagrangian a particular combination of curvature squared terms
$(R_{\mu\nu\gl\gs})^2-4(R_{\mu\nu})^2+R^2$ first considered in $D$
dimensions by Lovelock \cite{Lovelock}, with a coefficient of natural
magnitude $\sim 1$ in Planck units. 
The technical reason why I choose this combination
is that it leads to second order equations of motion. My
primary goal in this paper is to furnish an example of how things may go wrong,
and it is unlikely that any other combination would be better in this respect.

In this Einstein-Lovelock theory we will then consider the simplest ---
zero impact parameter --- collision case. Existence of a CTS for such
head-on collisions was shown long ago by Penrose \cite{Penrose}. We
will see that the curvature squared terms indeed become important and
significantly modify
the spacetime to the future of the transverse collision plane.

\begin{figure}[ht]
\includegraphics{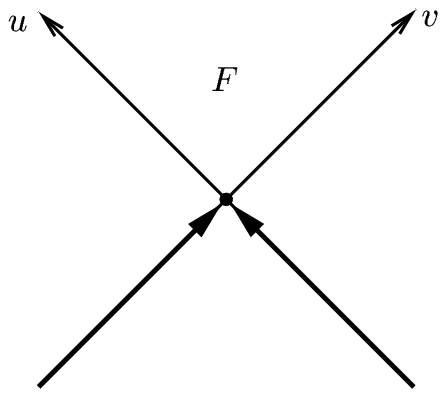}
\hspace{1.5cm}
\includegraphics{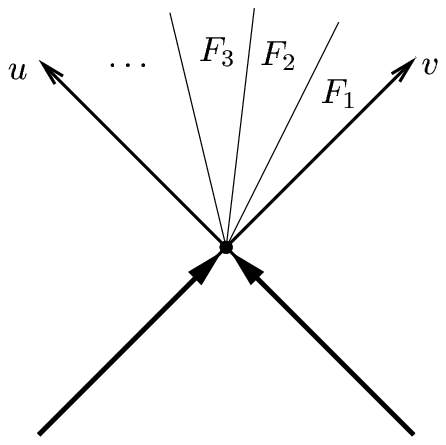}
\caption{The longitudinal slice of the collision spacetime in the pure
Einstein and the Einstein-Lovelock theories}
\label{fig0}
\end{figure}

The result of this analysis is summarized in Fig.\ \ref{fig0}. In 
the Einstein case the post-collision spacetime is weakly curved (at
large transverse radii).
The metric in the future wedge $F$ can be expanded in a power series
in the light-cone coordinates $u,v$, starting with a linear term
$\propto u+v$. The Riemann tensor has $\gd$-function
singularities on the null planes $u=0$ and $v=0$; the Ricci tensor is
of course everywhere zero. That curvature becomes large can be seen
from the
invariant $(R_{\mu\nu\gl\gs})^2$$\propto \gd(u)\gd(v)$. 

On the contrary, in the Einstein-Lovelock case the future wedge is split
into sectors $F_1,\ldots,$$F_n$, $n\ge 2$, separated by planes on
which the Riemann tensor has additional $\gd$-function singularities.
Moreover, the Ricci tensor is also singular on these planes, and in
general nonzero in between. In particular, the post-collision spacetime does not
satisfy Einstein's equations. An additional feature is that the number
of sectors $n$ is arbitrary; the solution is not unique.

The paper thus contains two {\bf main results} --- a general observation
(curvature becomes large) and a specific example (analysis in the
Einstein-Lovelock theory) --- which show that classical Einstein
gravity may not be applicable to study black hole production in
transplanckian collisions. If so, the geometric cross section
estimate \rf{geom} becomes much less certain, having lost its most
serious supporting argument.

The {\bf plan of the paper} is as follows. 
In Sec.~II we will review the CTS argument. Readers familiar with it
may go directly to Sec.~III, where we will discuss the danger posed by regions of high
curvature and see that curvature becomes large along the transverse
plane at the collision moment. This conclusion is reached after the
finite width of the shocks due to classical and quantum smearing is
taken into account.

The remaining part of the paper deals with the Einstein-Lovelock theory.
In Sec.~IV we will review the Lovelock Lagrangian and its higher order
generalizations. I will also comment on the history of this Lagrangian and
its alleged connection to string theory.

In Sec.~V we will write down a metric ansatz for the head-on collision,
in a plane-wave approximation. Then we will derive equations of motion, in a
small-gradient approximation, in the region of longitudinal
light-cone coordinates $|u|,|v|\lesssim 1$. We will also see that the cubic and
higher order Lovelock Lagrangians do not contribute in this
approximation.

In Sec.~VI we will solve the resulting hyperbolic Monge-Amp\`ere
equation. First we will see that no usual-sense (smooth) solution exist, even
when we consider the smeared out shocks as initial conditions. Then, in
the limit of infinitely thin shocks we will find an infinite class
of generalized (weak) solutions. We will discuss what this non-uniqueness
may mean, and how it can possibly be resolved.

In Sec.~VII we will check that the small-gradient approximation made
in Sec.~V was justified. We will then use the found solutions of the
Monge-Amp\`ere equation to write down approximate spacetime metrics.
I will explain the differences between the
Einstein and Einstein-Lovelock post-collision spacetimes, and their
negative implications for the CTS argument. I will conclude in Sec.\ VIII with a summary of the obtained results.

{\bf Notation.} In the remainder of the paper we work in flat
spacetime of $D\ge 5$ dimensions (although we quote some $D=4$ results
for comparison). This is legitimate, since the relevant gravitational
dynamics
happens in a region of $\sim \text{TeV}^{-1}$ size, which is much
smaller than the size of extra dimensions. 
We use Planck units,
setting $8\pi G=1$ in the $D$-dimensional Einstein-Hilbert action.
Greek indices $\mu,\nu,\ldots$ run through all $D$ coordinates, Latin
$i,j,\ldots$ only through $D-2$ transverse directions. Spacetime
signature is $(+,-,-,\ldots)$, and the curvature tensor definitions
are $R^\mu_{\ \nu\ga\gb}=\partial_\ga \Gamma^\mu_{\nu\gb}-\ldots$ and 
$R_{\mu\nu}=R^\ga_{\ \mu\ga\nu}$ (`Landau-Lifshitz timelike conventions').

\section{Closed Trapped Surface argument}

The basic ingredient in existing discussions of black hole formation
in particle collisions is the gravitational field of a fast point particle
moving along a straight line. In the limit of infinite boost
$\gamma=E/m\to\infty$ and fixed energy $E$, this field takes the form
of a shock wave
\begin{equation}
ds^2 = d\tu\, d\tv - \Phi(\tx)\, \delta(\tu)\, d\tu^2 - d\tx^{2}.
\label{AS}
\end{equation} 
Here $\tu=\Tt-\tz$, $\tv=\Tt+\tz$,
the particle is moving in the positive $\tz$ direction, and $\tx$
denotes the $D-2$ transverse coordinates.

Einstein's equations with the lightlike source
\be
T_{\bar{u}\bar{u}}=E\,\gd(\bar{u})\,\gd^{D-2}(\bar{x})
\ee
give the following equation for the shock wave profile $\Phi(\tx)$:
\begin{equation}
-\frac 12 \nabla^2 \Phi = E\, \delta^{D-2}(\tx).
\end{equation}
Thus we get\footnote{This result can also be derived by boosting the
  static $D$-dimensional Schwarzschild solution and simultaneously
  scaling down the rest mass to keep the energy constant. This was the
  method originally used by  Aichelburg and Sexl in $D=4$
  [\onlinecite{AS}].} ($\tr=|\tx|$; $\Omega_{D-3}$ is the volume
of the unit $D-3$ sphere)
\begin{eqnarray}
\Phi&=& -\frac E\pi \ln \tr \qquad (D=4),\label{four}\\
\Phi&=& \frac{2E}{\Omega_{D-3}(D-4)\,\tr^{D-4}}\qquad(D\ge5).\label{D}
\end{eqnarray}

Form \rf{AS} of the metric is unsuitable for analyzing the behavior of
geodesics crossing the shock at $\tu=0$, which is necessary for
understanding the causal structure. For this we must switch to
coordinates in which the metric is continuous. This is accomplished by
the discontinuous coordinate transformation [\onlinecite{D'Eath},
  \onlinecite{DH}, \onlinecite{EG}]
\begin{eqnarray}
\tu &=& u,\\
\tv &=& v+\Phi\theta(u) + \frac{u \theta(u) (\nabla\Phi)^2}{4},\\
\tx^i&=& x^i + \frac{u}{2} \nabla_i \Phi(x)\theta(u) 
\end{eqnarray}
(where $\theta$ is the Heaviside step function). In the new
coordinates the metric becomes
\begin{eqnarray}
 ds^2 &=& du\,dv - 
	H_{ik}H_{jk}\, dx^i dx^j,		\label{cont}	\\
  H_{ij} &=& \delta_{ij} + \frac  12\nabla_i\nabla_j \Phi(x)\,u\theta(u),
\end{eqnarray}
and both geodesics and their tangents are continuous across the shock.
Introducing polar coordinates in the transverse plane, this metric
can be written as (see \cite{D'Eath} for $D=4$)
\be
\label{mpolar}
ds^2=du\,dv-\Bigl[1+\frac{(D-3)E}{\Omega_{D-3}r^{D-2}}u\theta(u)\Bigr]^2dr^2-
\Bigl[1-\frac{E}{\Omega_{D-3}r^{D-2}}u\theta(u)\Bigr]^2r^2d\Omega^2.
\ee

To discuss a collision, we add a second fast particle of the same energy
moving opposite to the first one in the negative $z$ direction at a
transverse distance (impact parameter) $b$.
Assuming that the particles pass each other at $u=v=0$, the combined
gravitational field outside the wedge $u,v>0$ can be obtained by
superposing the metric \rf{cont} with its mirror image:
\begin{eqnarray}
 ds^2 &=& du\,dv - 
	(H_{ik}H_{jk}+\tilde{H}_{ik}\tilde{H}_{jk}-\gd_{ij})\, dx^i dx^j,		\label{coll}	\\
  \tilde{H}_{ij} &=& \delta_{ij} + \frac  12\nabla_i\nabla_j \Phi\bigl(x-(b,0,\ldots,0)\bigr)\,v\theta(v).
\end{eqnarray}
Such superposition is legal, because the excised region $u,v>0$ (wedge
$F$ in Fig.\ \ref{fig0}) is
precisely the future light cone of the collision plane of the
shocks. By causality, outside this region the shocks will not be able to influence
each other.

The metric for $u,v>0$ must be found by solving the characteristic
initial value problem for Einstein's equations. This task is
complicated by the fact that the shocks after passing each other will
focus and develop regions of high curvature. Shock propagation through
these regions depends on their detailed structure \cite{D'Eath}.

Nevertheless, we are interested whether this complicated dynamics will
lead to black hole formation. Since the complete metric is unknown, we
must resort to indirect arguments.
The idea [\onlinecite{Penrose}, \onlinecite{EG}] is to look for a closed trapped surface
(CTS) in the part \rf{coll} of the spacetime that we do know. CTS is
defined as a closed $(D-2)$-surface whose area decreases locally when
propagated along the outer null normals. Equivalently, the outer
null normals of such surface have positive convergence \cite{HE}.

Existence of a CTS in a spacetime solving vacuum Einstein's equations
implies presence of a singularity in the future. Assuming
Cosmic Censorship, this
singularity must be hidden behind a horizon, and we may conclude that
a black hole will form. Moreover, the black hole horizon must lie outside the CTS. Using this
information, one can get estimates of the horizon area and, via the
Area Theorem, of the mass of the formed black hole.

\begin{figure}[ht]
\includegraphics{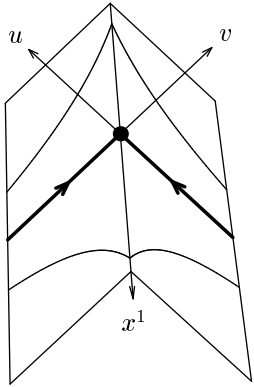}
\caption{CTS \rf{b=0}. The trajectories of the colliding particles and
  the null planes containing the CTS are also shown. All the
  transverse directions but $x^1$ are suppressed.}
\label{fig1}
\end{figure}

For the head-on collision ($b=0$) a CTS is easy to find
[\onlinecite{Penrose}, \onlinecite{EG}]. It lies in the union of the
shock planes $u=0$ and $v=0$ and,
when transformed to the $\tu,\tv,\tx$ coordinates, consists of two
flat $(D-2)$-dimensional disks of radii
\be
\label{r0}
r_0=(E/\Omega_{D-3})^{1/(D-3)}
\ee
centered at the collision point.
In the $u,v,x$ coordinates the CTS is glued out of two halves
described by 
\be\label{b=0}
\begin{aligned}
&u=0,\qquad v=\Phi(r_0)-\Phi(r),\qquad r\le r_0;\\
&v=0,\qquad u=\Phi(r_0)-\Phi(r),\qquad r\le r_0.
\end{aligned}
\ee
It is symmetric under rotations of the transverse directions and the
reflection $z\to -z$ (see Fig.\ \ref{fig1}). Strictly speaking, this surface is
a {\it marginal} CTS, which means
that the outer null normals have {\it zero} convergence. (Such a surface
is also called {\it apparent horizon}.) However,
moving a small distance inside one can find a true CTS with negative
convergence.

For nonzero impact parameters ($b>0$) the search of trapped surfaces
was carried out analytically for $D=4$ by Eardley and Giddings
\cite{EG} and numerically for $5\le D\le 11$ by Yoshino and Nambu
\cite{YN}. In every $D$ existence of a marginal CTS was established
for impact parameters
\be
b\le c_D\,
R_{\text{Schw}}(E)=c_D\Bigl(\frac{2E}{(D-2)\Omega_{D-2}}\Bigr)^{1/(D-3)},
\ee
where $c_D\approx 1.5$ is a numerical constant weakly depending on $D$.
This result of classical general relativity if the most serious
argument in favor of the geometric cross section for the black hole
production rate. 

\section{High curvature at $u=v=0$}

The CTS argument of the previous section was valid for the Einstein
gravity with its vacuum equation of motion $R_{\mu\nu}=0$. This
equation is used crucially in the Raychaudhuri equation describing how
the convergence $\theta$ evolves along the congruence of null normals
$\ell^\mu$:
\be
\label{Ray}
\frac {d\theta}{d\Lambda}=R_{\mu\nu}\ell^\mu\ell^\nu+\frac 12
\theta^2+2(\text{shear})^2.
\ee
For Ricci-flat spacetimes the first term vanishes. One then concludes
that once $\theta$ goes positive, it grows monotonically and becomes
infinite at finite affine distance $\Lambda$. This argument is the
basis of the proof of singularity theorems \cite{HE}.

Now suppose that long before the singularity forms, the evolving surface passes
through a region of high curvature. If higher order curvature terms
are present in the effective gravitational Lagrangian, they will
become important and modify the field equations at this point. In general,
$R_{\mu\nu}$ will no longer vanish, and one cannot make any
conclusions about the dynamics of $\theta$ across the region of high
curvature.

From the effective field theory point of view, it seems likely that all
interactions allowed by symmetry should become significant in a theory
at its natural scale (which is the $D$-dimensional Planck scale in our
case). This is the main reason to believe that the gravitational
Lagrangian will contain higher curvature terms,
with coefficients of natural magnitude $\sim 1$.
 There are also other hints pointing in the same direction.
For instance, since the Einstein gravity is non-renormalizable (see,
e.g., review \cite{deser}), higher curvature counterterms should be included
to cancel infinities in loop diagrams. In string theory, which
is a finite theory of quantum gravity \cite{books}, a particular sequence of higher
curvature terms with calculable coefficients appears in the effective
action (the $\ga'$-expansion).\footnote{For completeness, we have to mention that Loop Quantum Gravity \cite{lqg}
  attempts to quantize general relativity non-perturbatively, starting
  from the pure Einstein-Hilbert action. 
  Since the program is unfinished, and in particular the classical
  limit is poorly understood, it is hard to say what form the
  gravitational field dynamics will eventually take. It seems
  likely that the {\it effective} action, as long as this concept is
  applicable, will have to contain higher curvature terms even in this
  theory.}

It should be noted that singularity theorems of general relativity can
also be formulated and proved in presence of matter sources satisfying
various positivity assumptions, such as the null energy condition
$T_{\mu\nu}\ell^\mu\ell^\nu\ge 0$. While such conditions are natural
for classical matter, they can be violated by the renormalized
energy-momentum tensor of quantum fields \cite{BD}. So it is unlikely that any
useful generalization of singularity theorems exists when quantum
effects are taken into account.

These considerations make it clear that appearance of high curvature
regions is problematic for the whole CTS argument. Eardley and
Giddings tried to address this problem. In particular they noticed 
that the CTS they constructed lies in the planes of shocks.
They proposed to bring it out of these planes by propagating it a
small affine distance to the future along the outer null geodesics.
This way, they said, one can get a CTS lying everywhere in the region
of small curvature. The conclusion was that quantum gravity effects
are unlikely to modify the result: the black hole will still form. 

I believe that this argument as it stands is incomplete: it misses 
the fact that curvature may become large on the part of the trapped
surface where it crosses the transverse collision plane.

First of all, the only nonzero components of
the Riemann tensor of spacetime \rf{AS} are
\be
\label{Riem}
R_{uiuj}=\frac 12 \nabla_i\nabla_j\Phi(x)\,\gd(u).
\ee
All curvature invariants (contractions of this tensor with itself and
the metric) vanish identically. Thus a shock plane by itself 
is {\it not} a region of high curvature.\footnote{This also follows
  from the fact that the Aichelburg-Sexl wave can be obtained by boosting from manifestly low-curvature regions far
away from a static $D$-dimensional black hole.}

However, a problematic region does appear when we add a second shock. 
It is located at $u=v=0$, where the shocks collide. At this moment we
can form a nonvanishing curvature invariant of the
collision spacetime \rf{coll}:
\be
\label{blowup}
R_{\mu\nu\gl\gs}R^{\mu\nu\gl\gs}\propto \sum_{i,j}R_{uiuj}R_{vivj}\propto
\gd(u)\,\gd(v).
\ee
This equation seems to suggest that curvature becomes large (and even infinite) all
along the transverse plane $u=v=0$. However, it would be incorrect to
jump to this conclusion too soon. The reason is that infinitely thin
shocks are an idealization; in reality the shocks will have a finite
width $w$.

There are essentially two reasons for $w>0$. The first,
purely classical, reason is that in practice $\gamma=E/m$
is large but not infinite. Because of this, the shocks have width
\be
w_\text{class}\sim r/\gamma,
\ee
depending on the transverse distance $r$. This width comes out
naturally when deriving the shock wave by boosting the static black hole
spacetime [\onlinecite{Pirani}, \onlinecite{D'Eath}].

The second reason has quantum nature. In relativistic quantum theory
coordinates of a particle at rest cannot be measured more precisely than its
Compton wavelength $\sim 1/m$ (see, e.g., \cite{BLP}) . For an ultrarelativistic particle, this
limit becomes $\sim 1/E$. Thus the point particle picture used in the
CTS argument should not be taken literally. In reality the particles
should be thought to have finite size $\sim 1/E$, which translates to
a contribution to the shock width
\be
w_\text{quan}\sim 1/E.
\ee

We are interested in curvature at transverse radii $r\sim r_0$ relevant
for the trapped surface formation and evolution. In practice, 
for such $r$ we will always have $w_\text{class}\ll w_\text{quan}$, because the masses of
colliding particles are small compared to 1 TeV, and thus $\gamma$ is very
large. (E.g. $\gamma\sim 10^6$ at the LHC, using current-quark masses
$m_{u,d}\lesssim 10 \text{ MeV}$.) Thus we can estimate
\be
w\sim w_\text{quan}\sim 1/E.
\ee

Because of nonzero width, $\gd$-functions in \rf{Riem},
\rf{blowup} will be smeared out over an interval of length $\Delta u\sim
w$. Since the integral has to remain unity, the maximal attained
value will be $\sim 1/w\sim E$. Using this value in \rf{blowup},
we find
\be
\label{radius}
\max\limits_{r=r_0}\, (R_{\mu\nu\gl\gs})^2=4(D-2)(D-3)\,\Omega_{D-3}^\frac{2}{D-3}
\,E^{\frac{2(D-4)}{D-3}}.
\ee
This formula involves a {\it positive} power of $E$, and a large prefactor 
(increasing from $\sim 300$ to $\sim 670$ for $D=5\ldots 11$). 

We see that even after finite shock width is taken into account,
curvature still becomes large in the region of transverse collision plane
relevant for the black hole formation.
This means that corrections to the Einstein gravity will become
important at this moment. Their subsequent effect on dynamics is hard to predict in
general. It is not {\it a priori} excluded that this effect will be
transient, localized in a small shock-interaction region to the 
future of $u=v=0$, after
which the gravitational field will return to its Einsteinian value. In
this case we could push the CTS through this small problematic region,
similarly to Eardley and Giddings's proposal, and get to small
curvature values where the Einstein gravity would again be applicable.
(In the original proposal [\onlinecite{EG}, p.~6]
  the part of the trapped surface near $u=v=0$ was to be left fixed.)
However, I think that such localization is unlikely to happen. It is much more
probable that the corrections will be significant 
up to the values $u,v\sim 1$, and even further. This would mean
that the Einstein gravity alone essentially loses its predictive power
in the future of the collision plane (a possibility also mentioned by Kancheli \cite{Kan}).

In the remaining part of the paper I will give an example of how this scenario may be realized, using a particular modification
of the Einstein gravity by curvature squared terms.

\section{Einstein-Lovelock gravity}
We consider a modification of the $D$-dimensional Einstein gravity by
curvature squared terms described by the Lagrangian
\def\mL{\mathcal{L}}
\begin{eqnarray}
\mathcal{L}&=&-\frac 12 R\sqrt{-g}+\kappa\, \mathcal{L}^{(2)},\label{ELlagr}\\
\mL^{(2)}&=&(R_{\mu\nu\gl\gs}R^{\mu\nu\gl\gs}-4R_{\mu\nu}R^{\mu\nu}+R^2)\sqrt{-g}. \label{quad}
\end{eqnarray}
The new coupling $\kappa$ is assumed to be $\sim 1$ on the
grounds of naturalness. Lagrangian \rf{quad} has an interesting property first
discovered by Lovelock \cite{Lovelock}: it is the only combination of
curvature squared terms leading to second order
equations of motion for the metric. (A generic combination would produce fourth order equations.)

For this reason theory \rf{ELlagr} has been often used in attempts to
understand how higher curvature corrections modify the behavior of
pure Einstein gravity. Thorough reviews of the existing literature 
can be found in \cite{Myers} (applications to black holes) and \cite{DM}
(Kaluza-Klein scenarios; braneworlds). Recently, theory \rf{ELlagr} was
even discussed in connection with black holes produced in particle collisions:
the authors of \cite{gbbh} argued that by studying their Hawking
evaporation spectra one can measure $\kappa$.
However, in this paper we are interested how the Lovelock term manifests itself {\it
  before} rather than after the black hole formation, and whether it
may in fact preclude this formation.

Actually, Lovelock \cite{Lovelock} has discussed a whole sequence of
higher order Lagrangians
\be
\mL^{(n)}=\frac{(2n)!}{2^n} R_{[\nu_1 \nu_2}{}^{\nu_1 \nu_2} R_{\nu_3
    \nu_4}{}^{\nu_3 \nu_4}\ldots 
R_{\nu_{2n-1} \nu_{2n}]}{}^{\nu_{2n-1} \nu_{2n}}\sqrt{-g}.
\label{Ln}
\ee
This gives $R\sqrt{-g}$ for $n=1$ and reduces to \rf{quad} for $n=2$.

In general, $\mL^{(n)}$ vanishes identically for $D\le 2n-1$ and is a
total derivative for $D=2n$ (the generalized Gauss-Bonnet theorem
\cite{EGH}). For $D\ge 2n+1$, $\mL^{(n)}$ leads to second order equations of
motion. The corresponding variations were also found by Lovelock
\cite{Lovelock}:
\begin{align}
&\gd\Bigl(\int \mL^{(n)}\,d^Dx\Bigr)=\int G^{(n)}_{\gl\gs}\gd
g^{\gl\gs}\sqrt{-g}\,d^Dx,\\
\label{Gn}
& G^{(n)}_{\gl\gs}=-\frac{(2n+1)!}{2^{n+1}}g_{\gl[\gs} R_{\nu_1
    \nu_2}{}^{\nu_1 \nu_2} 
\ldots R_{\nu_{2n-1} \nu_{2n}]}{}^{\nu_{2n-1} \nu_{2n}}.
\end{align}
For $n=1$ this coincides with the Einstein tensor.

Explicit expressions for $\mL^{(n)}$ and $G^{(n)}_{\gl\gs}$ ($n\le 4$)
with terms of equivalent tensor index structure collected as in
\rf{quad} can be found in \cite{Briggs}. We won't need them, since in
practice for $n\ge 2$ it is much easier to work with \rf{Ln}
and \rf{Gn}.

Sometimes in the literature the theory \rf{quad} in $D=5$ is
attributed to Cornelius Lanczos, citing his papers \cite{Lan}.
In fact, however, Lanczos never went beyond $D=4$.\footnote{In 1932 he did
not even include $(R_{\mu\nu\gl\gs})^2$ in the Lagrangian; in 1938 he
proved that \rf{quad} is a topological invariant density in $D=4$. I
would
like to thank N.~Deruelle, J.~Madore and J.~Zanelli for the
interesting discussions
of this historical matter.}    
The name `Gauss-Bonnet' often associated with gravity theories 
based on Lagrangians \rf{Ln} 
is also an example of unfortunate terminology, since in $D\ge 2n+1$ these
`dimensionally continued' Gauss-Bonnet densities are no longer associated
with topological invariants (which is precisely why they become
interesting from the dynamical point of view). In this paper we will
use the term {\it Einstein-Lovelock gravity} for the theories based on
\rf{quad}, \rf{Ln}.

It is often incorrectly stated that $\mL^{(2)}$ is the only combination of curvature
squared corrections to the Einstein gravity consistent with string
theory. This claim is based on a (correct) result of Zwiebach \cite{Zwiebach},
who showed that \rf{ELlagr} is the only
curvature squared gravity theory in which 
the quadratic term in the expansion of the action around flat space is
the same as in the pure Einstein gravity. In particular, the graviton
is the only particle in the perturbative spectrum, and unphysical
ghost poles usually associated with curvature squared terms (see,
e.g., \cite{Stelle})
are absent. However, one should remember that only on-shell effective
action can be defined in string theory. This is especially clear when
this action is computed from the S-matrix, but is also true for the
sigma model $\beta$-function method (see, e.g., \cite{MT}). Thus there
always remains field-redefinition freedom, which can be used for
example to change the coefficients in front of or completely remove 
$(R_{\mu\nu})^2$ and $R^2$ terms in \rf{quad}. All these
actions would be equally consistent with the string S-matrix and, thus,
with string theory \cite{DR}. In a sense, string theory does not have
much to say about off-shell dynamics of quantum gravity.

I thus prefer to think of higher curvature theories like \rf{ELlagr} not
as fundamental microscopic theories, but as
effective field theories of gravity (see, e.g., recent discussion in \cite{burgess}), with the hope that they may capture some
aspects of gravitational field dynamics in presence of regions of high curvature.

\section{Metric ansatz and equations of motion}

We will now study the effect of the Lovelock term \rf{quad} on the collision spacetime at
$u,v>0$.
For simplicity, we will analyze the head-on
collision ($b=0$). In principle, the method could also be used to study
the nonzero impact parameter case, most certainly with similar
conclusions. Remember that we always assume $D\ge 5$.

Our collisions are transplanckian: $E\gg 1$. It is this situation that
can lead to formation of a large classical black hole according to
the common lore [\onlinecite{BF}, \onlinecite{EG}, \onlinecite{Landsberg}]. For $E\sim 1$ we would be speaking
about Planck-size black holes, for which quantum gravity effects are
significant without doubt.

\begin{figure}[ht]
\includegraphics{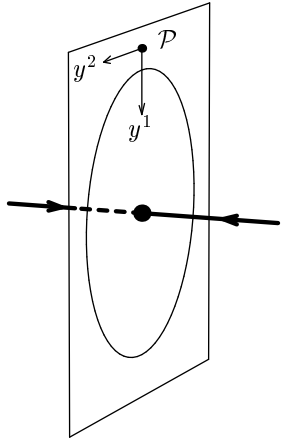}
\caption{The transverse collision plane. Shown are the trajectories of
  the colliding particles and the circular intersection with the CTS
  \rf{b=0}.}
\label{fig2}
\end{figure}

Our goal is to find the metric to the future of the collision
plane up to $u,v\lesssim 1$, at transverse radii $r\sim r_0$. 
Since $r\gg 1$, we can neglect transverse derivatives of the
metric: at $u,v\alt 1$ they
are suppressed by $1/r$ compared to the longitudinal ones [see \rf{coll}, \rf{mpolar}]. 
\def\P{\mathcal{P}}
Neglecting transverse derivatives means that we approximate the metric
near a given point $\P$ in the
collision plane located at $r \sim r_0$ by a plane wave collision metric (see \cite{Griff})
 \be
\begin{split}
\label{ansatz}
ds^2=e^{L(u,v)}du\,dv-&\bigl[1+A(u,v)\bigr]^2(dy^1)^2
-\bigl[1+B(u,v)\bigr]^2\sum_{i=2}^{D-2}(dy^i)^2.
\end{split}
\ee
Here $y^i$ are Cartesian coordinates in the transverse plane near
$\P$, with the $y^1$ axis pointing in the
direction of the collision point, and $y^2,\ldots y^{D-2}$ in the
$D-3$ orthogonal directions (see Fig.~\ref{fig2}). 
 
The initial conditions to be imposed on $L,A,B$
outside the future wedge $F(u,v>0)$ 
are obtained by matching \rf{ansatz} with
\rf{coll}, \rf{mpolar}. 
We find ($u,v\notin F$) 
\be
\label{init}
\begin{aligned}
L(u,v)&=0,\\
A(u,v)&=\eps\, \bigl[u\theta(u)+v\theta(v)\bigr],\\
B(u,v)&=-\frac{\eps}{D-3} \, \bigl[u\theta(u)+v\theta(v)\bigr],
\end{aligned}
\ee
where
\be
\label{eps}
\eps=\frac {(D-3)E}{\Omega_{D-3} r^{D-2}}\sim E^{-1/(D-3)} \ll 1.
\ee

\def\Ai{A^{(i)}}\def\Aj{A^{(j)}}\def\Ak{A^{(k)}}
To derive the equations of motion, it is convenient to work with a more
general metric\footnote{In the rest of this section we suppress Einstein's convention about
summing in repeated indices, and indicate all necessary summations explicitly.}
\be
\label{ans}
ds^2=e^{L(u,v)}du\,dv-\sum_{i=1}^{D-2}\bigl[\Ai(u,v)\bigr]^2(dy^i)^2.
\ee
The nonzero components of the Riemann tensor are ($\Ai_u,L_{uv},\ldots$ are partial derivatives)
\be
\label{R}
\begin{aligned}
R_{uvuv}&=\frac 12\,e^L\, L_{uv},\\
R_{uiui}&=[\Ai_{uu}-\Ai_u L_u]\,\Ai,\\
R_{uivi}&=\Ai_{uv}\,\Ai,\\
R_{vivi}&=[\Ai_{vv}-\Ai_v L_v]\,\Ai,\\
R_{ijij}&=-2\,e^{-L}[\Ai_u\Aj_v+\Ai_v\Aj_u]\,\Ai\Aj.
\end{aligned}
\ee

We will now make an {\it a priori} assumption that
\be
\label{sg}
|\nabla L|,\  |\nabla \Ai|\ll 1 \qquad(u,v\alt 1),
\ee
which also implies
\be
\label{sg1}
L\approx 0,\quad \Ai\approx 1\qquad (u,v\alt 1).
\ee
 Validity of this {\it small-gradient approximation} will have to be
checked after a solution is found. For now we see that it is compatible with the initial conditions \rf{init}.

The approximate Riemann tensor is found by applying \rf{sg} and
\rf{sg1} in \rf{R}. We have:
\be
\label{Ra}
\begin{aligned}
R_{uvuv}&\approx\frac 12 L_{uv},\\
R_{uiui}&\approx\Ai_{uu},\\
R_{uivi}&\approx\Ai_{uv},\\
R_{vivi}&\approx\Ai_{vv},
\end{aligned}
\ee
all other components being $\approx 0$.

The vacuum equations of the Einstein-Lovelock gravity are
\be
\label{EL}
-\frac 12 G_{\gl\gs}+\kappa G^{(2)}_{\gl\gs}=0.
\ee
The $uv$ and $ii$ components of this tensor equation will give $D-1$
independent equations for $D-1$ functions $L$ and $\Ai$.
The nonzero components of $G_{\gl\gs}$ and $G^{(2)}_{\gl\gs}$ in the small-gradient approximation are easy to find using \rf{Ra}. For $G_{\gl\gs}$ we have:
\be
\label{G}
\begin{aligned}
G_{uv}&=\sum_i \Ai_{uv},\\
G_{uu}&=-\sum_i \Ai_{uu},\\
G_{vv}&=-\sum_i \Ai_{vv},\\
G_{ii}&=-2L_{uv}-4\sum_{k\ne i} \Ak_{uv} \qquad(1\le i\le D-2).
\end{aligned}
\ee
For $G^{(2)}_{\gl\gs}$ we use the general expression \rf{Gn} with
$n=2$. Because of the antisymmetrization in the r.h.s.\ of \rf{Gn}, only terms for which
$\gs$, $\nu_1,\ldots$, $\nu_{2n}$ are all different can contribute. 
Also, looking at \rf{Ra}, we
see that $u$ or $v$ must be present in each pair of indices
$(\nu_1,\nu_2),\ldots,$ $(\nu_{2n-1},\nu_{2n})$. Using such reasoning, 
it is easy to see that for $n=2$ the only nonzero
components are\footnote{Notice that for $n\ge 3$ the given conditions
  on indices are incompatible. This means that the higher order
  Lovelock tensors vanish in the small-gradient approximation.}
\be
\label{G2}
G^{(2)}_{ii}=16\sum_{\substack{j\ne k\\j,k\ne
    i}}\bigl[\Aj_{uv}\Ak_{uv}-\Aj_{uu}\Ak_{vv}\bigr]\qquad(1\le i\le
D-2).
\ee

We now go back to our original ansatz \rf{ansatz}, putting
\be
A^{(1)}=1+A(u,v),\qquad \Ai=1+B(u,v)\qquad(2\le i\le
D-2).
\ee
Substituting \rf{G} and \rf{G2} in \rf{EL}, we have three independent
equations for $L,A,B$. The simplest is the $uv$-equation, to which
only the Einstein tensor contributes:
\be
A_{uv}+(D-3)B_{uv}=0.\label{uv}
\ee
This equation is easy to solve. Namely, it implies that the relation
\be
B(u,v)=-\frac {A(u,v)}{D-3},\label{BArel}
\ee
satisfied by the initial data \rf{init}, will continue to hold for
$u,v>0$.

There remain two $ii$ equations ($i=1,2$). After $B$ is excluded
using \rf{BArel}, they can be brought to the form 
\begin{eqnarray}
&&\kappa
  \bigl[A_{uu}A_{vv}-(A_{uv})^2\bigr]+\frac{D-3}{16(D-4)}A_{uv}=0, \label{Aeq}\\
&&L_{uv}=A_{uv}.\label{easy}
\end{eqnarray}
After a function $A(u,v)$ satisfying \rf{Aeq} is found, it will be
easy to find $L(u,v)$ from \rf{easy}. Taking the initial conditions
\rf{init} into account, we will have 
\be
\label{L}
L(u,v)=A(u,v)-\eps\bigl[u\theta(u)+v\theta(v)\bigr].
\ee

Thus we reduced the problem to solving a single nonlinear
second order PDE \rf{Aeq}. In the pure Einstein theory ($\kappa=0$) the nonlinearity
disappears, and we see that the solution for $u,v>0$ is given by the
same formulas \rf{init} as the initial data. For $\kappa\sim 1$ this is
no longer a valid solution, since it has
\be
\label{sing}
 A_{uu}A_{vv}-A^2_{uv}=\eps^2\,\gd(u)\,\gd(v),
\ee
which becomes large at $u=v=0$ (namely, $\sim\eps ^2 E^2 \gg 1$, when 
the finite shock width $w\sim 1/E$ and the resulting smearing of the
$\gd$-functions are taken into account).

PDE \rf{Aeq} belongs to the well-known family of {\it hyperbolic
  Monge-Amp\`ere equations}. It may be noticed (see \cite{Nutku1}) that it can be derived
  from the action 
\begin{align}
S_{\text{MA}} &=\int du\,dv\Bigl\{\frac{\kappa}3\,\bigl(A_u^2
A_{vv}-2A_v A_u A_{uv}+A_v^2 A_{uu}\bigr) 
+ \frac{D-3}{16(D-4)}A_uA_v\Bigr\}\notag\\
&=\int
dt\,dz\,\Bigl\{\frac{\kappa}3\,\epsilon^{aa'}\epsilon^{bb'}\partial_a
A\,\partial_bA\,\partial_{a'b'}A
+\frac{D-3}{64(D-4)}\eta^{ab}\partial_aA\,\partial_bA\Bigr\},
\label{MAact}
\end{align}
which we should consider as a 2-dimensional truncation of the full
Einstein-Lovelock action relevant for our problem.
The first term in this action uses only the totally
antisymmetric tensor to form contractions. For this reason the
corresponding nonlinear term in Eq.~\rf{Aeq} is invariant with
respect to Euclidean rotations as well as Lorentz transformations. The linear wave
equation term in \rf{Aeq} is of course only Lorentz-invariant. 

\section{Solving the Monge-Amp\`ere equation}

\subsection{Smooth solutions}

 Hyperbolic Monge-Amp\`ere equations have exceptional
  character among the nonlinear second order hyperbolic
  PDEs:
  they can be reduced to a system of five first order characteristic
  ODEs, while in general (in two
  dimensions) one needs eight  [\onlinecite{CH}, p.~495]. However, in case of
  Eq.~\rf{Aeq}, because it has constant coefficients, this general
  theory is superseded by an even simpler method based on the Legendre
  transform.

Let us rewrite \rf{Aeq} as
\be
\label{A}
A_{uu}A_{vv}-A^2_{uv}+\gl A_{uv}=0,\qquad
\gl=\frac{D-3}{16(D-4)\kappa},
\ee
substitute
\begin{eqnarray}
&&A=\gl(\tilde{A}+uv),\\
&&\tilde{A}_{uu}\tilde{A}_{vv}-\tilde{A}^2_{uv}- \tilde{A}_{uv}=0,
\end{eqnarray}
and apply the Legendre transform [\onlinecite{CH}, p.~32]. This
results in a linear Poisson equation, which is readily solved. In this way
Ignatov and Poponin \cite{IP} obtained an exact solution of the
original Monge-Amp\`ere equation \rf{A} in an implicit form, depending on two
arbitrary functions $F(\xi)$ and $G(\eta)$:
\be\label{IP}
A(u,v)=\gl\bigl[F(\xi)+G(\eta)+F'(\xi) \,G'(\eta)\bigr],
\ee
where $\xi$ and $\eta$ must be found from
\be\label{trans}
\begin{aligned}
u&=\eta+F'(\xi),\\
v&=\xi+G'(\eta).
\end{aligned}
\ee

For vanishing at infinity small functions $F,G$ this solution
describes interaction of two colliding pulses of the $A$ field. The
pulses emerge from the interaction region with their shape unchanged,
resuming the motion along the pre-collision world lines.

In general, solution \rf{IP} is applicable if the transformation
$(\eta,\xi)\to(u,v)$ given by \rf{trans} is one-to-one. This means
that its Jacobian should not vanish:
\be
\label{J}
\frac{\partial(u,v)}{\partial(\eta,\xi)}=1-F''(\xi)\,G''(\eta)\ne 0.
\ee
Unfortunately, this condition is violated in our problem. Indeed, to
satisfy the initial conditions \rf{init}, we would have to take
\be
\label{F}
F(\xi)=G(\xi)=\eps\,\xi\,\theta(\xi).
\ee
Because of finite shock width $w\sim 1/E$ these initial data have to
be smeared out on this scale. Still this results in $F''(\xi)$ near
$\xi=0$ of the order $\sim\eps E \gg 1$.
So the Jacobian \rf{J}, being equal to unity away from the origin,
changes sign and becomes negative near $\xi=\eta=0$. 
The transformation \rf{trans} is thus not one-to-one:
there is a region of the $u,v$ plane which is covered 3 times, and where
the inverse transformation is multiple-valued. 

This analysis shows that for initial data \rf{init}, even
after smearing them out, the exact solution \rf{IP} in any case
cannot be used verbatim. A region of spacetime appears where this
solution is triple-valued, and we have to somehow choose between the
three branches. Actually, it turns out that no satisfactory choice is possible.
More precisely, any choice introduces discontinuities either in the
function $A$ itself (which is totally unacceptable) or in its
first derivatives (which brings back the problems which we temporarily
resolved by using smoothed out initial data). All of them seem to violate Eq.~\rf{A}. 

To get some insight about the source of these difficulties, I also studied the 
problem numerically. The conclusion of this study (which I am not
going to report here in detail) is that the solution develops singularities characterized by
discontinuous first partial derivatives. Evolution beyond these
singularities apparently cannot be described by formula \rf{IP}. A
different strategy is required, which I will now proceed to describe.

\subsection{Generalized solutions}

To find a solution, we will have to extend the class of admissible functions
$A$, allowing continuous functions with discontinuous first
derivatives. [Notice that the initial data \rf{init}
corresponding to infinitely thin shocks belong to precisely this
class.] Such generalized solutions, called {\it weak} in the theory of PDEs
[\onlinecite{CH}, p.~418, 486], are usually defined via integration by
parts. In our case weak solutions can be defined, because the
Monge-Amp\`ere operator can be written as a divergence
\be
\label{diverg}
A_{uu}A_{vv}-A^2_{uv}=\partial_v(A_vA_{uu})-\partial_u(A_vA_{uv}).
\ee
Thus Eq.~\rf{A} is equivalent to the conservation law
\be
\label{claw}
\partial_uJ_v+\partial_vJ_u=0,
\ee
where the current $J_a$ has components
\begin{align}
J_v&=-A_vA_{uv}+\frac{\gl}2 A_v,\\
J_u&=A_vA_{uu}+\frac{\gl}2 A_u.
\end{align}
For smooth functions $A$ the differential
form \rf{claw} of the conservation law is equivalent to the integral form:
\be
\label{ilaw}
\oint_\mathcal{C} \epsilon^{ab} J_a\, dx_b =0
\ee
for any closed curve $\mathcal{C}$. However, the latter condition does not
involve products of second derivatives of $A$ and can be unambiguously
checked even when the r.h.s.~of \rf{A} cannot be defined. I thus take
\rf{ilaw} as the definition of weak solutions of Eq.~\rf{A}. 
  
In practice, it is always sufficient to check \rf{ilaw} only for
infinitesimally small contours $\mathcal{C}$, and this only near special, most
dangerous, points. In the rest of the $u,v$ plane the differential
equation \rf{A} can still be used. Provided that the first partial
derivatives of $A$ are bounded, the nonsingular terms proportional to $\gl$ drop
out of \rf{ilaw} in the limit of small contours. The remaining rule for catching possible $\gd(u)\,\gd(v)$-type singularities as in
\rf{sing} takes the form
\be
\int\limits_{\sqrt{u^2+v^2}\le\epsilon}(A_{uu}A_{vv}-A^2_{uv})\,du\,dv=\oint\limits_{\sqrt{u^2+v^2}=\epsilon}
    A_v\,dA_u.\label{Stokes}
\ee
[This formula has in fact general validity and follows directly from \rf{diverg}.] If there is a $\gd$-function at the origin present in the integrand on
the l.h.s., we will be able to detect it by studying the limit of the r.h.s.
as $\epsilon\to 0$. The latter procedure does not involve products of
distributions and is unambiguous.

\def \a{\mathcal{A}}
For further discussion it is convenient to simplify \rf{A} by making
the substitution
\begin{eqnarray}
&&A= \mathcal{A}(u,v)+\frac \gl 2 uv,\label{AAA}\\
&& \mathcal{A}_{uu}\mathcal{A}_{vv}-\a^2_{uv}+\gl^2/4=0. \label{a'}
\end{eqnarray}
This PDE is hyperbolic due to positivity of $\gl^2/4$.

We will no longer attempt to use smeared out initial data, and
instead will try to find a solution directly in the limit
of infinitely thin shocks. Since in this limit $u=v=0$ is likely to be the most singular
point, we change to polar coordinates 
\begin{align}
u=\rho \sin \varphi,\qquad v=\rho \cos\varphi.
\end{align}
PDE~\rf{a'} becomes
\be
\frac
1{\rho^2}\bigl[\a_{\rho\rho}(\a_{\varphi\varphi}+\rho\,\a_\rho)-(\a_{\rho\varphi}-\a_\varphi/\rho)^2\bigr]+\gl^2/4=0.
\label{a}
\ee
The crucial fact (easy to guess if one remembers the well-known connection
between the homogeneous Monge-Amp\`ere equation and developable surfaces) is that
for functions $\a$ of the form $\rho f(\varphi)$ the first most singular
term in \rf{a} vanishes. To satisfy the full equation, we have to
include in $\a$
terms of order $\rho^2$. Thus we are led to the ansatz\footnote{It is possible to consider a more general ansatz, adding terms
  $\rho^n h_n(\varphi)$, $n\ge 3$. The functions $h_n$ would then be
  restricted by an infinite sequence of nonlinear ODEs, obtained
  similarly to \rf{a1} and \rf{a2} below. These higher order terms
  can be neglected compared to \rf{aa} in the limit of small $\rho$,
  and so we put them to zero for simplicity. In principle, however,
  the possibility of them being nonzero further aggravates the
  non-uniqueness issue to be discussed in Sec.~VI.C below.}:
\be
\label{aa}
\a=\rho f(\varphi)+\rho^2 g(\varphi).
\ee

\begin{figure}[ht]
\includegraphics{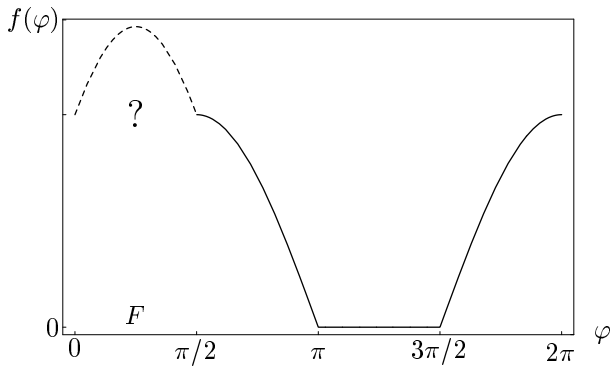}
\caption{The initial data for $f(\varphi)$. In the future wedge $F$
  (see Fig.\ \ref{fig0}) the
  function has to be determined (?). The dashed line
  denotes the Einsteinian solution \rf{fwrong}.}
\label{fig3}
\end{figure}

Compatibility with the initial conditions \rf{init} requires:
\be
\label{f}
\begin{aligned}
f(\varphi)&=\begin{cases} 
\eps \sin\varphi,&\varphi\in[\pi/2,\pi],\\
0,&\varphi\in[\pi,3\pi/2],\\
\eps \cos\varphi,&\varphi\in[3\pi/2,2\pi],
\end{cases}
\\
g(\varphi)&=-\frac\gl 4 \sin 2\varphi,\qquad \varphi\in[\pi/2,2\pi].
\end{aligned}
\ee
For $\varphi\in[0,\pi/2]$ the functions $f,g$ have to be found so that
PDE~\rf{a'} is satisfied (see Fig.~\ref{fig3}).

Checking PDE \rf{a'} at $u=v=0$ calls for the use of identity
\rf{Stokes}. An easy calculation shows that 
\be
\oint_{\rho=\epsilon}\a_v\,d\a_u=\frac
12\int_0^{2\pi}\bigl(f^2-f'{}^2\bigr)\,d\varphi+O(\epsilon).
\ee
Taking \rf{f} into account, we get a condition for the absence of $\gd(u)\gd(v)$ in \rf{a'}:
\be
\label{nodelta}
\int_0^{\pi/2}\bigl(f'{}^2-f^2\bigr)\,d\varphi=0.
\ee
As an example when $\gd(u)\gd(v)$ does appear, we can take the
Einsteinian solution
$\a=\eps\bigl[u\theta(u)+v\theta(v)\bigr]$, which would correspond to
\be
\label{fwrong}
f(\varphi)=\eps(\sin\varphi+\cos\varphi),\qquad \varphi\in[0,\pi/2].
\ee
And indeed we readily see that \rf{nodelta} is violated.

At $\rho>0$ the solution can be understood in the usual sense, and the
use of \rf{Stokes} is unnecessary. We simply plug the ansatz \rf{aa}
in Eq.~\rf{a}. Equating the coefficients before $\rho^{-1}$ and
$\rho^0$ to zero, we get two ODEs:
\begin{align}
&(f''+f)\,g=0, \label{a1}\\
&2g\,g''-g'{}^2+4g^2+\gl^2/4=0,\label{a2}
\end{align}
which have to be solved with the following from \rf{f} boundary
conditions
\begin{align}
&f(0)=f(\pi/2)=\eps,\label{b1}\\
&g(0)=g(\pi/2)=0.\label{b2}
\end{align}
At the first glance we have a problem here, since ODE~\rf{a1} implies
that either $g=0$ or 
\be
\label{osc}
f''+f=0.
\ee
The first opportunity seems to be excluded by \rf{a2}, while the only
solution of \rf{osc} with the boundary conditions \rf{b1} is
\rf{fwrong}, which does not satisfy the no-delta condition \rf{nodelta}.

The way out of this impasse is to notice that although $g$ is indeed
not allowed to be zero on an {\it interval}, it can still vanish at
{\it isolated points} inside $(0,\pi/2)$. At these points the
derivative $f'(\varphi)$ can have jump discontinuities. If on the rest of
the interval \rf{osc} holds, ODE \rf{a1} will still be satisfied.

The general solution of ODE \rf{a2} depends on two constants
$C,\bar{\varphi}$ and has the form
\be
\label{solg}
g(\varphi)=\pm \frac \gl 4
\Bigl\{\sqrt{C^2+1}\cos\bigl[2(\varphi-\bar{\varphi})\bigr]-C\Bigr\}. 
\ee
At points $\varphi_*$ where $g(\varphi_*)=0$ we have, in agreement with
\rf{a2},
\be
g'(\varphi_*)=\pm\gl/2.
\ee
The constants $C,\bar{\varphi}$ and the overall sign in \rf{solg} can be
chosen independently on the two sides of $\varphi_*$, provided that both
choices are compatible with $g(\varphi_*)=0$. In particular $g'(\varphi_*)$
is allowed to flip sign at $\varphi_*$. Being multiplied by vanishing $g$, the
arising $\gd$-function in $g''$ does not violate \rf{a2}.

Taking all these observations into account, we can write down the
following {\bf class of solutions} to the boundary value problem
\rf{a1}-\rf{b2}:
\begin{itemize}
\item
Functions $f,g$ are continuous on $[0,\pi/2]$.
\item
 The $f$ satisfies \rf{osc}
everywhere on $[0,\pi/2]$ except at some chosen points
\be
\label{interm}
0<\varphi_1<\ldots<\varphi_N<\pi/2,
\ee
where $f'(\varphi)$ will have jump discontinuities. 
\item
The $g$ vanishes
at $\varphi_1,\ldots,\varphi_N$, and is given by \rf{solg} on each subinterval
$[\varphi_\ga,\varphi_{\ga+1}]$ ($\varphi_0=0$,
$\varphi_{N+1}=\pi/2$ are the end points) with
\begin{align}
\bar{\varphi}&=\frac 12(\varphi_{\ga+1}+\varphi_\ga),\\
C&=\cot(\varphi_{\ga+1}-\varphi_\ga),
\end{align}
and an arbitrary overall sign $\pm$, which can be chosen independently on each subinterval.
\end{itemize}

\begin{figure}[ht]
\includegraphics{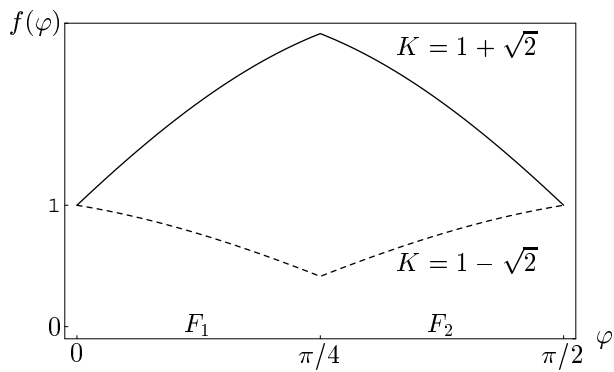}
\caption{Solution \rf{ex1}. The future wedge $F$ is split into two equal
  sectors $F_{1,2}$ (see Fig.\ \ref{fig0}). The leading behavior of
  $A(u,v)$ in these sectors is given by $u+Kv$ and $v+Ku$.}
\label{fig4}
\end{figure}

\begin{figure}[ht]
\includegraphics{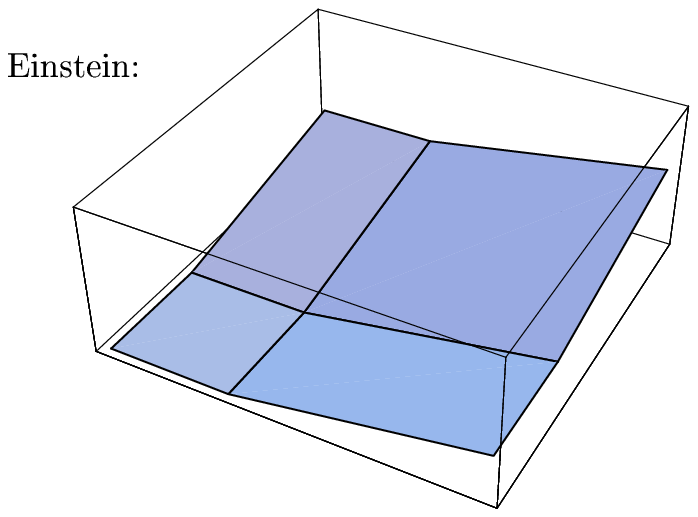}

\includegraphics{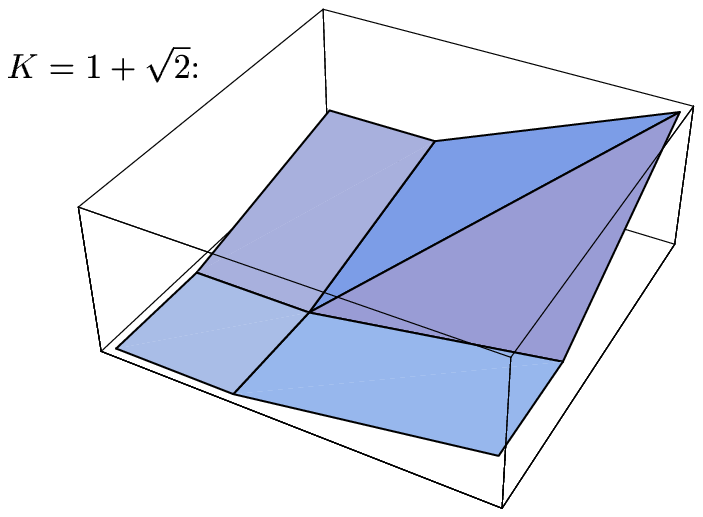}
\hspace{1cm}
\includegraphics{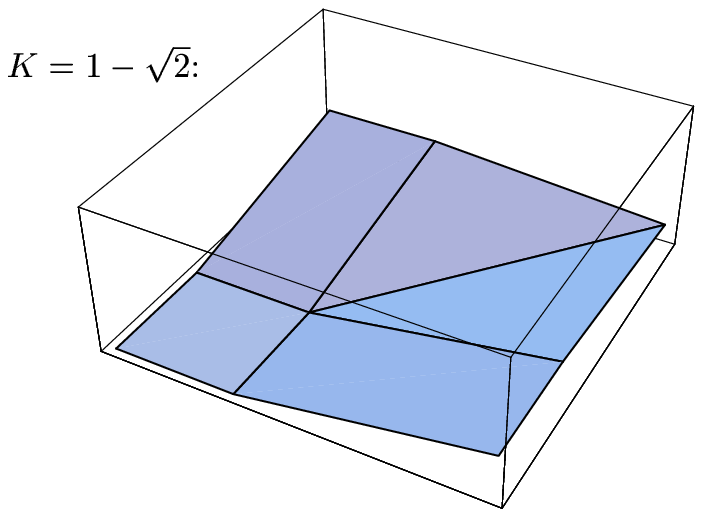}
\caption{The leading behavior of $A(u,v)$ near $u=v=0$ for the pure
  Einstein case \rf{fwrong} and for the Einstein-Lovelock solution \rf{ex1}. The
  time flow is from the left to the right.}
\label{fig5}
\end{figure}

Within this class it becomes possible to satisfy the no-delta condition
\rf{nodelta}. The simplest and most symmetric example can be
constructed using one intermediate point $\varphi_1=\pi/4$ (see Fig.~\ref{fig4}). From
continuity and \rf{osc} we must have
\be
\label{ex1}
f(\varphi)=\eps\times\begin{cases}
\cos\varphi+K\sin\varphi,& \varphi\in[0,\pi/4],\\
\sin\varphi+K\cos\varphi,& \varphi\in[\pi/4,\pi/2].
\end{cases}
\ee
The constant $K$ is determined from \rf{nodelta}. We find that two
values are allowed:
\be
K=1\pm\sqrt{2}.
\ee
For $g(\varphi)$ we have:
\be
g(\varphi)=\pm\frac\gl 4\times\begin{cases}
\sqrt{2}\cos(2\varphi-\pi/4)-1,& \varphi\in[0,\pi/4],\\
\sqrt{2}\cos(2\varphi-3\pi/4)-1,& \varphi\in[\pi/4,\pi/2].
\end{cases}
\ee
Near the point $u=v=0$ the leading $\rho f(\varphi)$ 
behavior of these solutions (as of
all the solutions in the class described above) is piecewise linear in
terms of the $u,v$ coordinates,
with the $\rho^2g(\varphi)$ term being a small correction (see Fig.\
\ref{fig5}).

A more general example, which we will need in a discussion below,
utilizes two symmetric intermediate points $\varphi_1=\phi$,
$\varphi_2=\pi/2-\phi$, where $\phi\in[0,\pi/4]$. The function $f$ is
given by
 \be
\label{ex2}
f(\varphi)=\eps\times\begin{cases}
\cos\varphi+K\sin\varphi,&\varphi\in[0,\phi],\\
\displaystyle\frac{\cos\phi+K\sin\phi}{\cos\phi+\sin\phi}(\cos\varphi+\sin\varphi),&
\varphi\in[\phi,\pi/2-\phi],\\
\sin\varphi+K\cos\varphi,&\varphi\in[\pi/2-\phi,\pi/2],
\end{cases}
\ee
where the $K$ is fixed by \rf{nodelta} to be
\be
\label{K}
K=1\pm\sqrt{1+\cot\phi}
\ee
The allowed functions $g$ can be written down according to the rules
given above. For $\phi=\pi/4$ this example reduces to the previous
one.

Taking even more general configurations of intermediate points and imposing
the no-delta condition \rf{nodelta}, we can get infinitely many solutions of PDE~\rf{a'}
having the form of the ansatz \rf{aa} and satisfying the initial conditions \rf{f}.
Solutions of the original Monge-Amp\`ere equation \rf{A} are then
given by Eq.~\rf{AAA}.

\subsection{Discussion of non-uniqueness}

Let us repeat the logic of the preceding discussion. In Sec.~VI.A we
convinced ourselves that solutions of the Monge-Amp\`ere equation
\rf{A} with the smeared out initial conditions \rf{init}
run into singularities, characterized by discontinuous first partial
derivatives.
The very meaning of a solution in presence of such discontinuities
needs to be reconsidered. For this reason in Sec.~VI.B we introduced
the definition of weak solutions, using a standard procedure of integration by parts.\footnote{Notice that for smooth
functions the concepts of usual and weak solutions are equivalent. It
is only in presence of discontinuous first derivatives that the usual
notion is not applicable, and we have to use the new definition.}
Working in the limit of infinitely thin shocks, we
found a whole class of weak solutions with the required initial
conditions. It is of course satisfying to find that solutions
according to the new definition {\it exist}, but how should we
interpret their non-uniqueness?

In my opinion, there are two alternative possibilities. First, it is
possible that this non-uniqueness can be resolved by performing a
detailed study of PDE \rf{A} with smeared out initial data at
small but finite shock width $w\sim 1/E$. As we have just discussed, the
singularities will still appear. But perhaps by studying their
approach one may discover a preferred way to continue the solution
past them. This would require techniques beyond what we use in this
paper. However, we would like to point out that even if a `true'
unique solution is found in this way, it will still have to satisfy
our weak solution criteria from Sec.~VI.B. Because of this it seems
likely (although cannot be guaranteed) that this `true' solution  will
be close to one of the weak solutions we found in Sec.~VI.B.

The second, more intriguing, possibility is that the encountered
non-uniqueness is fundamental. Then its interpretation can be most
naturally given in quantum theory. It would simply mean that beyond
the shock collision the gravitational field wavefunction ceases to be
concentrated on a single classical configuration. Instead, it spreads out
over all allowed solutions of the equations of motion. In this case we
would have to sum over all solutions, using $\exp(iS)$, where $S$ is
the on-shell value of the action, as the weight.

It should be noted that the problem of non-uniqueness of weak
solutions is well known in the theory of PDEs. It is present for
instance in the theory of first order quasilinear hyperbolic systems,
where the weak solutions in question are shock waves. There the
non-uniqueness is resolved by imposing the physically motivated `entropy
condition' \cite{Lax}.

One can try to do something similar in spirit for our PDE \rf{A}. For
example, one may try to identify higher order currents that are conserved for usual
smooth solutions, and impose their conservation as an additional
constraint on weak solutions.
In fact, the starting point of our discussion of weak solutions was to
  rewrite the Monge-Amp\`ere equation \rf{A} itself as the current
  conservation condition \rf{claw}. Going a step further, let us introduce a symmetric
  2-tensor $T_{ab}$ with components
\be
\label{EM}
\begin{aligned}
T_{uu}&=(\nabla A)^2 A_{uu},\\
T_{vv}&=(\nabla A)^2 A_{vv},\\
T_{uv}&=T_{vu}=-(\nabla A)^2 A_{uv}
\end{aligned}
\ee
We have 
\begin{align}
\partial_v T_{uu}+\partial_u T_{vu}&=2 A_v(A_{uu}A_{vv}-A^2_{uv}),\label{emc1}\\
\partial_v T_{uv}+\partial_u T_{vv}&=-2 A_u(A_{uu}A_{vv}-A^2_{uv}).\label{emc2}
\end{align}
Actually, the appearance of the Monge-Amp\`ere operator in the
r.h.s.~is not surprising: $T_{ab}$ is nothing but the energy-momentum
(EM) tensor of the homogeneous Monge-Amp\`ere equation,
  Eq.~\rf{A} with $\gl=0$. This equation
corresponds to the first term in the action
\rf{MAact}. To define the symmetric EM tensor, we covariantize this term:
\be
\begin{split}
\int
d^2x\,&\epsilon^{aa'}\epsilon^{bb'}\partial_a
A\,\partial_bA\,\partial_{a'b'}A
\longrightarrow \int
d^2x\sqrt{g}\,E^{aa'}E^{bb'}\partial_a
A\,\partial_bA\,\nabla_{a'}\partial_{b'}A 
\end{split}
\ee
and differentiate with respect to the auxiliary 2-dimensional metric
$g_{ab}$. The result of this computation is \rf{EM} (up to a factor of
$3/2$).

The EM tensor of the full equation \rf{A} differs from \rf{EM} by a
trivial harmless piece associated with the quadratic term $(\partial
A)^2$ in \rf{MAact}. In imposing
the EM conservation at the singular point $u=v=0$ that piece is irrelevant.
Transforming the conservation laws \rf{emc1}, \rf{emc2} to their
integral form, we conclude that the EM tensor will be conserved if
\be
\oint_{\rho=\epsilon}(\nabla A)^2 dA_{u},\quad 
\oint_{\rho=\epsilon}(\nabla A)^2 dA_{v}\to 0\qquad(\epsilon\to 0).
\ee
Applied to the ansatz \rf{aa}, this gives two conditions on 
$f(\varphi)$, conveniently written down as one complex
condition:
\be
\label{cons}
\oint d\varphi\,e^{i\varphi} (f'{}^2+f^2)\cdot(f+f'')=0.
\ee
When this condition is imposed, the class of weak solutions from
Sec.~VI.B is reduced, but not eliminated. The simplest solution
\rf{ex1} does not satisfy \rf{cons} and would not be allowed. However,
already among the one-parameter family \rf{ex2} there is a solution
[corresponding to the numerical value $\phi\approx0.322$ and the minus
sign in \rf{K}] which passes the
new criterion. Introducing more intermediate points, we will still
have an infinitude of solutions satisfying both the no-delta
condition \rf{nodelta} and the EM conservation \rf{cons}.

It seems likely that a whole hierarchy of currents conserved on the
smooth solutions of PDE \rf{A} can be found. In fact, it is possible
\cite{nutku2} to map the Monge-Amp\`ere equation \rf{a'} to the
Born-Infeld equation
\be
\label{BI}
(1+\Phi_x^2)\Phi_{tt}-2\Phi_t\Phi_x\Phi_{tx}-(1-\Phi_t^2)\Phi_{xx}=0,
\ee
or to the first order system describing dynamics of the Chaplygin gas
\be
\label{Ch}
\begin{cases}
P_t=PP_x+Q^{-3}Q_x,\\
Q_t=(PQ)_x.
\end{cases}
\ee
Both \rf{BI} and \rf{Ch} are known to possess infinitely many
conserved quantities \cite{nutku3}. Whether translating them back to the
Monge-Amp\`ere variables and imposing on the weak solutions is
physically motivated, or can help resolve the non-uniqueness, is an
open question.

This is as far as our explorations have taken us in the attempts to
resolve the issue of non-uniqueness. Unfortunately, the problem still
remains. In particular, at present we are unable to decide which one
of the two possibilities outlined at the beginning of this subsection
is realized. However, both of them allow for comparison with the pure
Einstein case, to which we now proceed.

\section{Approximate Post-collision Metrics --- Comparison}

In Sec.~V, when deriving the equations of motion,
we introduced a small-gradient assumption \rf{sg}, which for
our particular ansatz \rf{ansatz} takes the form
\be\label{SG}
|\nabla A|,\ |\nabla B|,\ |\nabla L| \ll 1\qquad(u,v\alt 1).
\ee
This was compatible with the initial data, but could not be taken for
granted for the full solution. So before we use the solutions found in
Sec.~VI.B, we have to check \rf{SG}. 

The functions $B$ and $L$ are
trivially related to $A$ by \rf{BArel} and \rf{L}. Thus it suffices
to check \rf{SG} only for $A(u,v)$, which is given by \rf{AAA} and
\rf{aa}. To estimate the gradient of \rf{aa}, we compute
\begin{align}
\nabla_u\bigl[\rho f(\varphi)\bigr]&=f(\varphi)\sin\varphi+
f'(\phi)\cos\varphi,\\
\nabla_u\bigl[\rho^2 g(\varphi)\bigr]&=\rho\,[2 g(\varphi)\sin\varphi+
  g'(\varphi)\cos\varphi\bigr]
\end{align}
Analogous formulas with $\sin\varphi$ and $\cos\varphi$ interchanged
are valid for $\nabla_v$.  
Since $f(\varphi)$ will typically be of order $\eps\ll 1$
[as in \rf{ex1} and \rf{ex2}], the $\rho f(\varphi)$ term passes the
small-gradient check.
Further, the size of $g(\varphi)$ is set by the parameter $\gl$, which 
will be $\sim 1$ for $\kappa \sim 1$. It follows that the $\rho^2
g(\varphi)$ term, as well as the term $\gl u v/2$ in \rf{AAA},
will also pass the check, provided that we restrict $u,v$ to a somewhat
smaller region
\be
\label{final}
u,v\alt \rho_{\max}\ll 1,
\ee
where the precise value of $\rho_{\max}$ has to be chosen depending on
the required accuracy. This value is insensitive to $E$, which
determines the shock width $w\sim 1/E$. This means that
the found solutions do take us out of the shock interaction region
before the small-gradient approximation breaks down.

The plane wave collision metrics \rf{ansatz} corresponding to the
solutions of Sec.~VI.B approximate the spacetime near the
chosen point $\mathcal{P}$ in the transverse plane. They implicitly depend on
$\mathcal{P}$ through the parameter $\eps$. To assure the consistency of neglecting transverse
 derivatives, we have to assume that all other free parameters, such
 as the choice of intermediate points $\varphi_\ga$, depend weakly on
 or are independent of
$\mathcal{P}$. Then we 
can patch the solutions for different $\mathcal{P}$'s together. 
The resulting approximate metric has the form
 \be
\begin{split}
\label{af}
ds^2=e^{L(u,v,r)}du\,dv-&\bigl[1+A(u,v,r)\bigr]^2dr^2
-\bigl[1+B(u,v,r)\bigr]^2 d\Omega^2,
\end{split}
\ee 
where the dependence of $L,A,B$ on the transverse radius is now made
explicit. These are the metrics described in the Introduction. 
The final range of their validity is
\be
r \sim r_0,\quad u,v\alt \rho_{\max}.
\ee

In the pure Einstein case, the approximate post-collision metric at
$r\sim r_0$, $u,v\alt 1$ would
have the same form \rf{af}, except that the function $A$ has to be
found from the linear wave equation $A_{uv}=0$ (see
Sec.~V). Thus we have
\begin{align}
A_{\text{E}}(u,v,r)&=\eps(r)\cdot[u\theta(u)+v\theta(v)],\\
B_{\text{E}}(u,v,r)&=-\frac{A_{\text{E}}(u,v,r)}{D-3},\\
L_{\text{E}}(u,v,r)&=0.
\end{align}
Comparing the pure Einstein and the Einstein-Lovelock
spacetimes, we see that they are quite different. In the Einstein
case the Riemann tensor is concentrated on the null planes, while in
the Einstein-Lovelock case it has additional $\gd$-function
singularities of comparable strength on the planes $u/v=\tan\varphi_\ga$
corresponding to the intermediate points \rf{interm}. This
difference persists as far as we can trace the solutions, that is up
to $u,v\sim\rho_{\max}$.

It is instructive to see what happens with the dynamics of CTS in such
a spacetime. In Einstein gravity, once a CTS appeared, it can be
evolved into the future along its null normals. The CTS property is
preserved by such evolution because of the Raychaudhuri equation
\rf{Ray}. This fact crucially depends on the inequality
\be
\label{ineq}
R_{\mu\nu}\ell^\mu\ell^\nu\ge0
\ee
(for any lightlike $\ell^\mu$), which follows from Einstein's
equations when the matter sources are either absent or satisfy
appropriate positivity conditions \cite{HE}.

The Ricci tensor of spacetime \rf{af} 
is best written down in the same local coordinates $u,v,y^i$ that we
used in Eq.\ \rf{ansatz}. Neglecting
transverse derivatives, and using the small-gradient approximation, we have
\be
\label{Ricci}
\begin{aligned}
R_{uv}&\approx -L_{uv}=-A_{uv},\\
R_{y^1y^1}&\approx 4A_{uv},\\
R_{y^iy^i}&\approx4B_{uv}=-\frac {4A_{uv}}{D-3}\qquad(2\le i\le D-2)
\end{aligned}
\ee
[where we used \rf{BArel} and \rf{L} to express $B$ and $L$ via $A$].
We see that in the Einstein-Lovelock case $R_{\mu\nu}\ne 0$ in the
future wedge. Most notably, $R_{\mu\nu}$ will have singularities on the planes
$u/v=\tan\varphi_\ga$,
with the leading behavior $\propto\rho\,\gd(\varphi)$. The magnitudes
and signs of these singularities depend on the particular choice of
solution. 

To see what happens with condition \rf{ineq}, let us pick a light-like
vector $\ell^\mu=$$(\ell^u,\ell^v,\ell^i)$. Since the metric
components do not differ much from their Minkowskian values (only
their second derivatives do), we have
\be
\label{null}
\ell^u\ell^v-(\ell^i)^2\approx 0.
\ee
Using \rf{Ricci} and \rf{null}, it is easy to compute
\be
R_{\mu\nu}\ell^\mu\ell^\nu=2A_{uv}\Bigl[(\ell^1)^2-\frac{D-1}{D-3}\sum_{i=2}^{D-2}(\ell^i)^2\Bigr].
\ee
The signs of both factors in this expression are undetermined, and so 
Ricci tensor \rf{Ricci} will generally violate condition \rf{ineq}.

We thus see that in the considered example the higher
curvature terms significantly modify the post-collision spacetime. It
appears that the pure Einstein gravity and the CTS argument based on it cannot be trusted in the future of the collision plane.

\section{Final remarks}

In this paper I questioned the use of classical Einstein gravity
to analyze black hole production processes in transplanckian particle
collisions. My basic argument was very simple: I
pointed out that curvature, as measured by the curvature invariant
$(R_{\mu\nu\gl\gs})^2$, becomes large on the transverse plane $u=v=0$
in the classical collision spacetime, in the region relevant for the horizon formation.

In principle, I could have stopped right here, since 
this fact alone already implies that quantum
corrections cannot be ignored.\footnote{Perhaps some readers will find the following additional explanation
  useful. 
It is sometimes stated that quantum gravity corrections in
transplanckian collisions should be small because the Schwarzschild
radius $R_{\text{Schw}}(E)\gg 1$ for $E\gg 1$. This would indeed be
true were $R_{\text{Schw}}(E)$ the only relevant length scale in the
problem. But my analysis shows that it is not: the other relevant scale is set by the curvature
radius near $u=v=0$ at $r\sim R_{\text{Schw}}(E)$, which according
to \rf{radius} is actually much smaller than the Planck length.}
However, I decided to go a bit further
and give a concrete example of how taking quantum corrections into
account may dramatically change the picture.

Needless to say, the precise form of quantum corrections to classical
general relativity remains so far unknown. However, there is a common
opinion (supported by string theory) that at least to some extent
these corrections may be represented by adding to the effective
action of gravity a sequence of higher curvature terms.

In this paper we considered a particular example of such `higher
curvature gravity' --- the so-called Einstein-Lovelock theory. We re-analyzed the collision in this theory and
found that the spacetime does deviate significantly from the pure
Einstein case. Thus, the initial conclusions are supported.

Of course, once it is found that a certain combination of higher
curvature terms strongly modifies the Einstein gravity result,
Pandora's box is open, and there is no reason to expect that all other
higher curvature terms can be excluded from the analysis. Moreover,
since the width of the colliding shock waves is $\sim
E^{-1}\ll 1$, it seems
very likely that the higher the power of curvature, the more important
this term will become. (It was not so for the higher Lovelock
  Lagrangians solely because of antisymmetrization.) 

For this reason, the given analysis of the Einstein-Lovelock gravity case
should not perhaps be assigned importance beyond that of an example
supporting the main argument. (Nevertheless, it seems to be the first study of a
  {\it dynamical} process in this theory; as far as we know, all previous applications
  involved static solutions.) The real conclusion of this analysis and
the whole paper is that, at the present state of knowledge about
quantum gravity, no well-founded claims about transplanckian
collisions at small impact parameters can be made. Even if TeV-scale
gravity is realized in nature, the true story of such collisions 
will be much more complicated than a single
large black hole production with a subsequent Hawking evaporation.

\begin{acknowledgments}
I would like to thank J.~de Boer, L.~Cornalba, D.~Labutin,
 A.~Naqvi, A.~Polyakov, A.~Sinkovics, K.~Skenderis, M.~Taylor, and the
 anonimous referee for many useful suggestions. This work was
 supported by Stichting FOM.  
\end{acknowledgments}

\end{document}